\documentclass[11pt,twoside]{article}
\usepackage{graphicx,epsfig,natbib,epstopdf}
\usepackage{CS18}
\begin{document}
%
%
%
\title{The HST Treasury ``Advanced Spectral Library'' (ASTRAL) Programs}
%
%
\author{Kenneth G. Carpenter$^{1}$, Thomas R. Ayres$^{2}$, ASTRAL Science Team$^{3}$}
\affil{$^1$NASA's GSFC, Code 667, Greenbelt, MD USA 20771}
\affil{$^2$University of Colorado, CASA 389-UCB, Boulder, CO 80309 }
\affil{$^3$Variety of domestic and international institutions}
\begin{abstract}
%
%
The ``Advanced Spectral Library (ASTRAL)'' Project (PI = T. Ayres) consists of two Treasury Programs: the Cycle 18 ``Cool Stars'' (GO-12278) Program  and the Cycle 21 ``Hot Stars'' (GO-13346) Program.   The primary goal of these programs is to collect, for the use of the astronomical community over the coming decades, a definitive set of representative, high-resolution (R$\sim$30,000-100,000), high signal/noise (S/N$>$100) spectra, with full UV coverage ($\sim$1150 - 3100 \AA) of prototypical stars across the HR diagram, utilizing the high-performance Space Telescope Imaging Spectrograph (STIS). The Cycle 18 program obtained spectra of 8 F-M evolved late-type stars, while the Cycle 21 program is in the process of observing 21 early-type stars, which span a broad range of spectral types between early-O and early-A.   All of these data will be available from the HST archive and, in post-processed and merged form, at  http://casa.colorado.edu/$\sim$ayres/ASTRAL/. These data will enable investigations of a broad range of problems -- stellar, interstellar, and beyond -- for many years into the future.  We describe here the details of the observing programs, including the program targets and the observing strategies utilized to optimize the quality of the spectra, and present some illustrative examples of the on-going scientific analyses, including a study of the outer atmospheres and winds of the two evolved M stars in the sample and a first look at a ``high definition'' UV spectrum of a magnetic chemically peculiar ``Ap'' star.
\end{abstract}
%
%
%
%
%

\section{Introduction and Overview}

 Over the past three years, two of the largest-ever HST stellar spectroscopic Guest Observer programs have been undertaken.  The ``Advanced Spectral Library (ASTRAL)'' Project (PI = T. Ayres) consists of two Treasury Programs, the first in Cycle 18 on ``Cool Stars'' (GO-12278) and the second in the current Cycle 21 on ``Hot Stars'' (GO-13346).   These programs are designed to collect a definitive set of representative, high-resolution (R$\sim$30,000-100,000), high signal/noise (S/N$>$100) spectra, with full UV coverage ($\sim$1150 - 3100 \AA) of prototypical stars across the HR diagram.  The Space Telescope Imaging Spectrograph (STIS) is ideally suited for this purpose and is the sole instrument used for the acquisition of the program spectra. 
 
The 8 F-M evolved late-type stars observed in ASTRAL/Cool Stars include iconic objects like Betelgeuse, Procyon, and Pollux. In addition, this sample has been supplemented with HST  observations from a separate Chandra program, so that two main sequence stars ($\alpha$~Cen A \& B) are included in this definitive dataset.  

ASTRAL/Hot Stars is in the process of observing 21 early-type stars, which span a broad range of spectral types between early-O and early-A, including main sequence and evolved stars, fast and slow rotators, as well as chemically peculiar and magnetic objects.  The targets include equally iconic Sirius, Vega, and the classical wind source Zeta Puppis. 

All of these extremely high-quality STIS UV echelle spectra will be available from the HST archive.  High-level science products, produced by the ASTRAL Science Team using the best-available techniques to calibrate and merge the individual spectra into a definitive, single spectrum, can be found at  http://casa.colorado.edu/$\sim$ayres/ASTRAL/. 

In Section 2 we describe the details of the observing programs and spectra now available for community use.  In Section 3 we present some illustrative examples of the on-going scientific analyses by the ASTRAL Science Team, including a study of the outer atmospheres and winds of the two evolved M stars in the sample, the M3.4~III giant Gamma Cru and the M2~Iab supergiant Alpha Ori and a first look at a ``high definition'' UV spectrum of the magnetic chemically peculiar ``Ap'' star HR 465. These serve as examples of the kind of research that these data can enable, as the data will support  investigations of a broad range of problems -- stellar, interstellar, and beyond -- by many investigators, for many years into the future.

\section{Targets and Observing Strategies}

In this section, we describe the content of these programs and the observing strategies used to optimize the final, high-level science products being delivered.  

\subsection{ASTRAL-I: Cool Stars}

The Cool Stars Program was carried out in HST Cycle 18 from October 2010 through November 2011.  It was awarded 146 orbits to record eight bright late-type stars, including well-known stars like Betelgeuse and Procyon. Cool stars are faint in UV, requiring $\sim$20 orbits per target to achieve good S/N at full wavelength coverage with the STIS echelles. The resolution of the spectra vary from R = 30,000 to 110,000 depending on wavelength and target brightness, as the goal was to obtain the  highest resolution possible with reasonable exposure times in a given wavelength regime. This sample has been supplemented with observations of equivalent quality from a separate Chandra program, which had HST observations as a component,  so that two main sequence stars ($\alpha$~Cen A \& B) are included in this dataset, along with the 8 evolved stars.  

The Science Team includes many Co-Investigators, covering a variety of topical areas and providing complementary ground-based observations and relevant atomic physics information for the analysis of the spectra. The ASTRAL-I program was driven by a desire to enable the study of: 1) magnetic activity, which is most conspicuous at the highest energies (FUV and X-rays), 2) stellar mass loss, which is most prominently seen in the UV resonance lines of abundant elements such as H~I, O~I, C~II, Fe~II, and Mg~II, 3) the interstellar medium, also seen in the same UV resonance lines, and 4) the complex atmospheric dynamics exhibited by these stars, due to the mysterious chromospheric and coronal heating mechanism(s), seen in the FUV ``hot lines", such as Si~III, Si~IV, C~IV, etc.

The resulting spectra, in compressed/downsampled form to permit display in a single figure, are illustrated in Fig. 1. {\bf We note that these data are the spectral equivalent of the Hubble Imaging ``Deep Fields'', in that one can zoom in on the final, co-added spectrum repeatedly and reveal more and more features and information, due to their very high S/N and resolution}.

\begin{figure}
\includegraphics[scale=0.8]{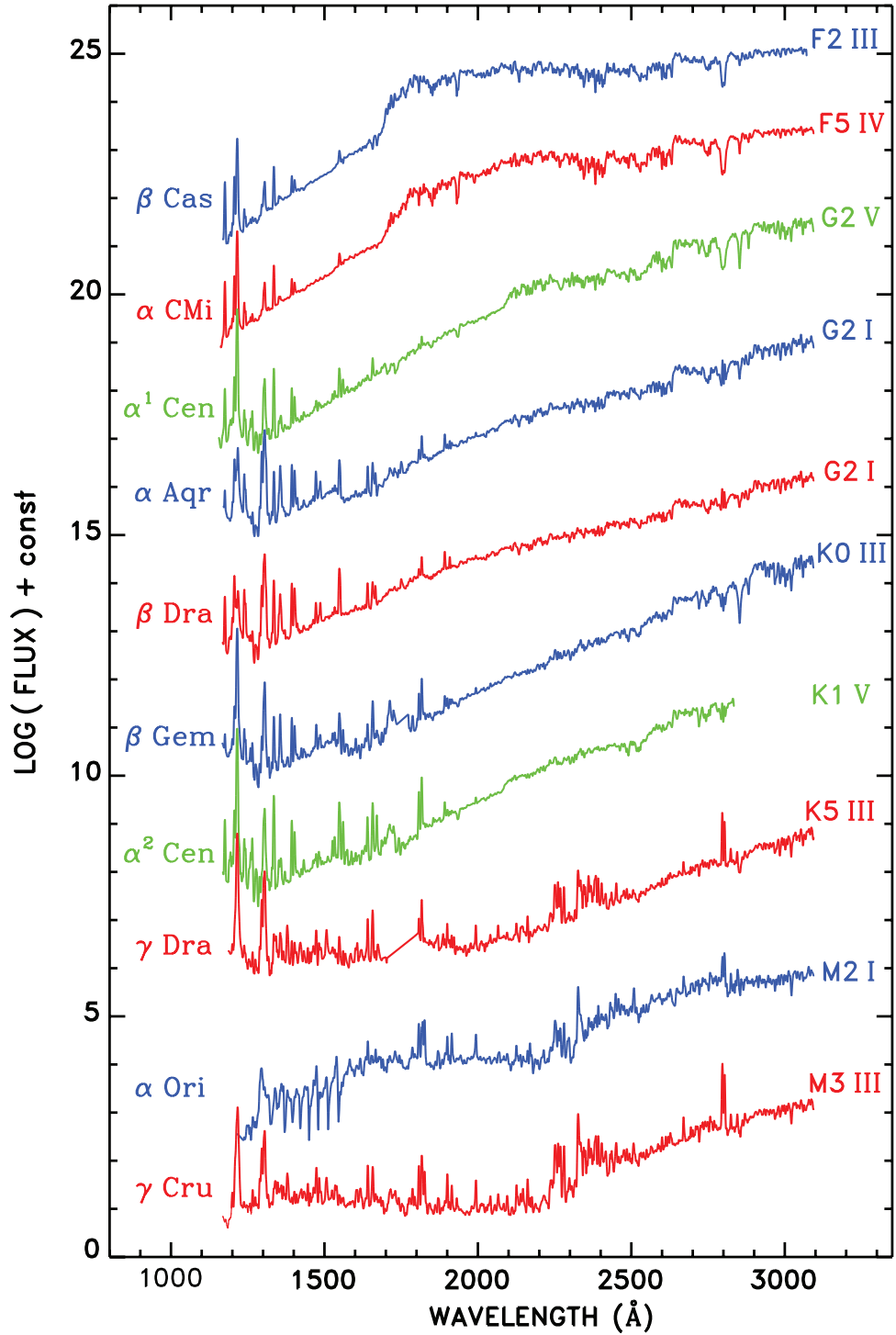}
\caption{STIS spectra smoothed to 1-\AA\ resolution. Red and blue: ASTRAL Cool Stars from Cycle 18; green: two solar-like dwarfs  ($\alpha$~Centauri A \& B) for comparison to the ASTRAL sample, which is mostly giants. Typical late-type stars show photospheric continuum longward of 2000 \AA\ (and Mg II $\lambda$2800 \AA\ emission), but shortward display numerous chromospheric emission lines. Strong 1300 - 1600 \AA\ absorptions in $\alpha$~Ori are cold circumstellar CO bands.}
\end{figure}

The Observing Strategy used to maximize the quality of these spectra, in terms of S/N and resolution, within the time available, is shown in Fig. 2. The primary guidelines used to formulate this strategy were to: 1) use highest-resolution mode/setting available and narrowest slit possible, within a reasonable total exposure time, 2) exposures were divided into subexposures to mitigate fixed pattern noise, and 3) at least one observation of each type was scheduled near a peak-up to ensure the best possible flux and wavelength calibrations.

Post-processing of the spectra corrects wavelength distortions identified in previous studies of STIS wavecal lamps, updated with new CALSTIS dispersion solutions from T. Ayres' ``Deep Lamp Project'' and special techniques were used to optimally splice and merge the orders of the echellegrams.

\begin{figure}
\includegraphics[scale=0.6]{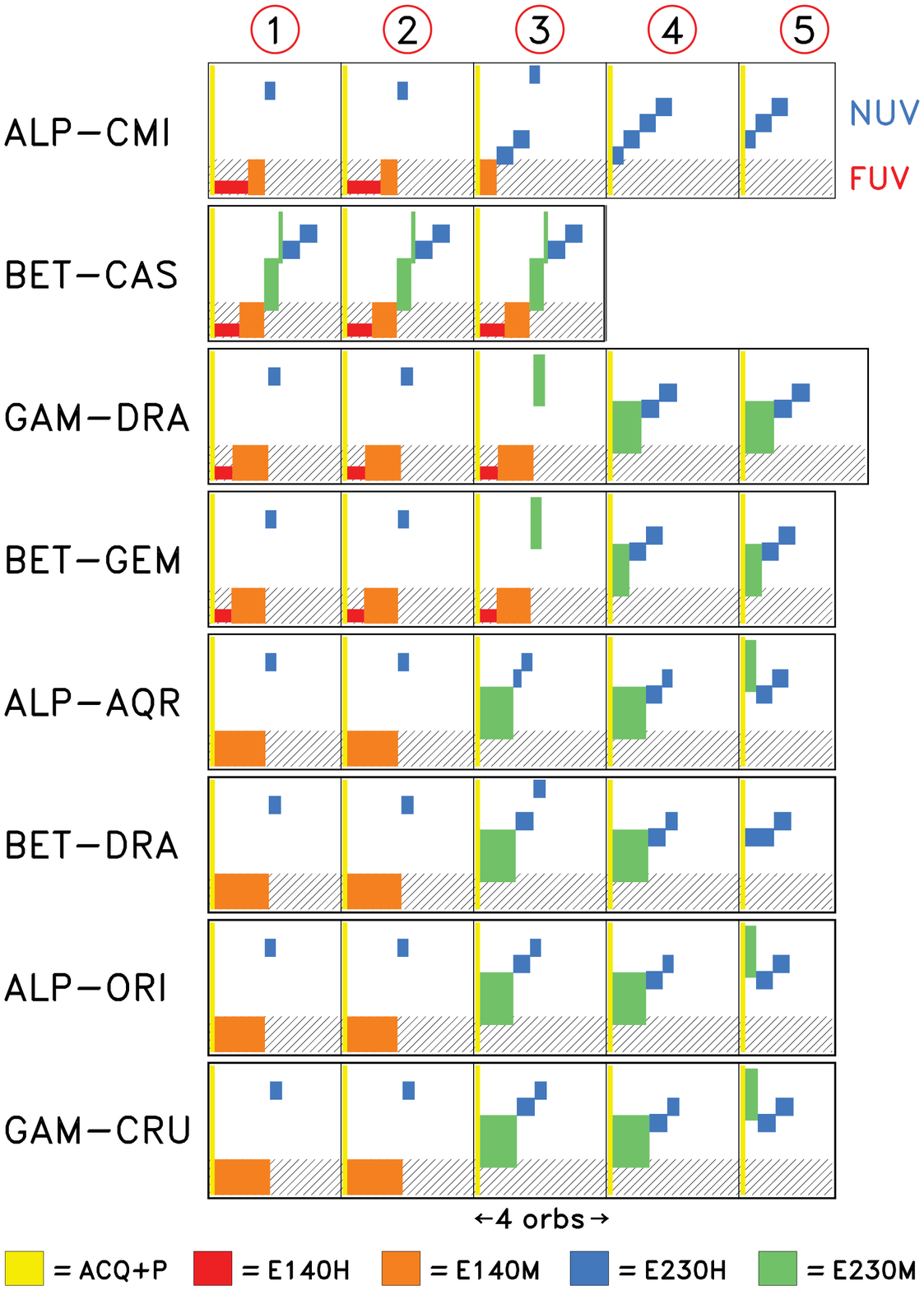}
\caption{Exposure sequences for ASTRAL-I. Each square box represents a 4-orbit visit (truncated boxes are 3). Colored rectangles indicate exposures in individual modes, according to legend at bottom. Horizontal width of symbol depicts total exposure duration (usually in multiple subexposures); vertical extent represents wavelength span of setting (y-axis: 1100-3200~\AA; hatched band is FUV interval; unshaded, NUV). Longer exposures were divided between two or more visits, to take advantage of slight non-repeatability of grating mode select mechanism (MSM), which places the echellegrams on different sets of pixels in independent visits, equivalent to an FP-split, suppressing fixed pattern noise. ``Empty'' space to right of each visit box represents unavoidable overheads, mainly Earth blocks, but also MSM movements, auto-wavecals, slit changes, and the like. Beta Cas can be observed in the Continuous Viewing Zone, so Earth occultations are avoided.}
\end{figure}

\subsection{ASTRAL-II: Hot Stars}

The Hot Stars Program focuses on the warm half of the H-R diagram.  It was awarded 230 orbits in the current Cycle 21, starting in August 2013 and continuing through most of 2014.  The Program observes 21 targets, including star-gazer favorites Vega, Sirius, and Regulus, and classic windy-star Zeta Puppis; normal and chemically peculiar stars, fast and slow rotators; magnetic and other exotica. Eleven targets have been fully observed to date.  The Program spectra are illustrated schematically in Fig. 3, which shows compressed ASTRAL spectra in color and ``place-holder'' IUE spectra in black for stars not yet observed in the ASTRAL-II program.

\begin{figure}
\plotone{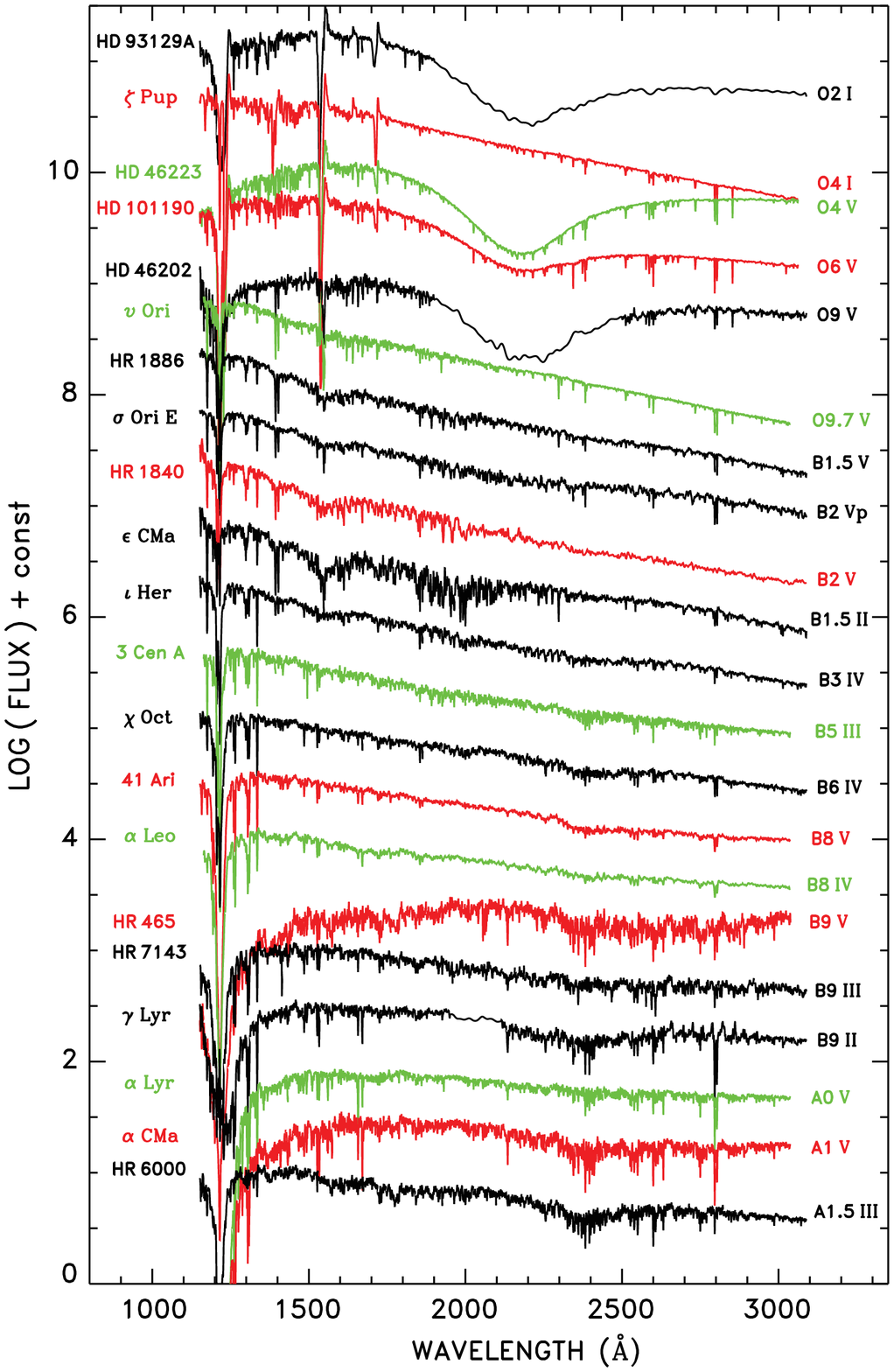}
\caption{ASTRAL-II SEDs. STIS spectra smoothed to 1-\AA\ resolution. Red and green: ASTRAL Hot Stars already observed in Cycle 21. Black is used for stars still to be observed - the data shown are from IUE. Note the strong interstellar absorption (2200 \AA\ ``bump'') in the spectra of several of the hottest stars. Strong metallic line-blanketing is seen at the shortest wavelengths of the cooler stars, especially in the Ap star HR~465, which has strong absorption throughout the UV - so much so that there is almost no true continuum point in this region.}
\end{figure}

We summarize in Fig. 4 the observing strategy used to optimize the signal-to-noise and resolution of each spectrum within the allocated observing time. We note in particular that this program was only made feasible due to the execution of a special GO calibration program, between the times of the ASTRAL-I and ASTRAL-II programs, to commission several previously ``available but unsupported'' neutral density slits on STIS.  The previously supported slits had a huge gap in attenuation.  The standard, clear slits transmit too much light and would have saturated the detectors, while the supported neutral density slits (ND=2 and ND=3) attenuate the light by 2 and 3 orders of magnitude (factors of 100x and 1000x), respectively, and  would have required too much observing time to be practical. T. Ayres in  GO Program 12567 (``Bridging STIS's Neutral Density Desert") therefore calibrated a set of ``available but unsupported'' ND (Neutral Density)-filtered long slits (31X0.05NDA/NDB/NDC) that can be used with the STIS echelles, and which provide intermediate attenuations between the standard (ND=0) spectroscopic slits (0.2X0.06, 0.2X0.09, or 0.1X0.03) and the (only two) supported ND slits: 0.2X0.05ND (ND=2) and 0.3X0.05ND (ND=3). These intermediate ND slits provide attenuations of ND=0.6, 1.0, and 1.4 (factors of 4x, 10x, and 25x). The use of these slits can be seen in Fig. 4.  In this figure, tall rectangles represent 96-minute HST orbits; 3-4  orbits constitute a ``visit'' (circled numerals). Colored rectangles indicate echelle settings; vertical size is wavelength coverage. The shorter rectangles are high-resolution settings, taller are medium-resolution settings, and horizontal size is exposure duration.  The lower hatched zone highlights FUV (1150 - 1700 \AA); upper clear band is NUV (1700 - 3100 \AA).  The guidelines for the observing strategies were similar to those used for ASTRAL-I: 1) use highest-resolution mode/setting and narrowest spectral slit possible within reasonable exposure times, 2) break up exposures to mitigate fixed pattern noise, and 3) schedule at least one observation of each type near peak-up.  In addition, special neutral density slits, as noted above were used for brighter objects.

\begin{figure}
\plotone{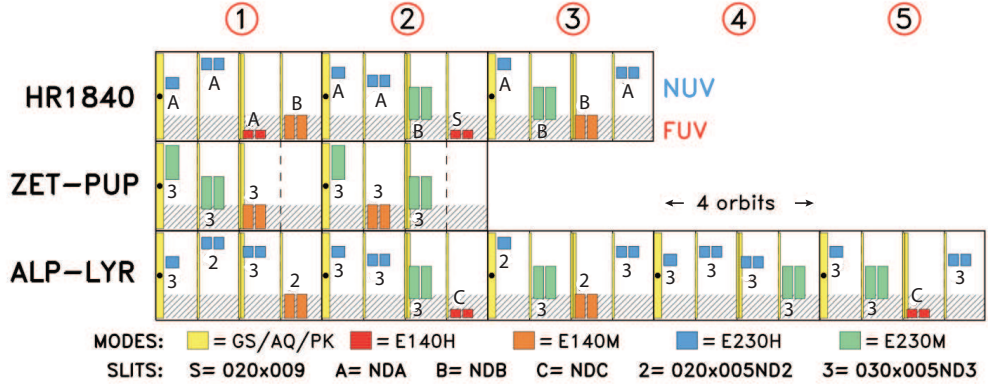}
\caption{Observing sequences for ASTRAL-II. Only part of each orbit is useful owing to Earth occultations.  Optimum echelle slits, target-dependent, are marked (see legend): special neutral density slits are used for UV-brighter objects. The upper panel illustrates 2 M (medium resolution) settings (orange \& green) plus 5 H (high resolution; red \& blue), for sharp-lined stars, requiring three 4-orbit visits.  The middle panel shows observations using exclusively M (3 settings), for broad-lined objects, requiring two 3-orbit visits. The bottom panel shows the special case for Vega, which requires five 4-orbit visits.  Non-redundant exposure sequences were used to optimize wavelength accuracy, preserve radiometric precision, and suppress detector fixed pattern noise.}
\end{figure}

\section{A Sampling of Early Science Results}

%
Here we present a couple of examples of some of the early science results from the ASTRAL-I and ASTRAL-II Programs.

\subsection{Science Example \# 1: Diagnostics of Outflowing Winds in M-Stars}

One of the studies being performed as part of the ASTRAL-I program is an investigation of the dynamics and wind of the coolest stars. As part of this, we have used observations of Fe~II lines of different strength to map the wind acceleration in the two M-stars in the ASTRAL-I sample, $\alpha$~Ori (M2~Iab) and $\gamma$~Cru (M3.4~III). The Fe~II profiles vary with the intrinsic strength of the lines, as the photons detected at earth originate higher in the atmosphere of the star for lines of higher opacity.  Therefore, as we view lines of increasing line strength, two things happen: 1) the chromospheric emission lines increase in width due to opacity-broadening, and 2) the overlying wind absorptions become stronger and shift to bluer wavelengths as the wind velocity increases with height.  This behavior is shown in Fig. 5, which shows that the $\alpha$~Ori wind acceleration appears small or non-existent compared to that seen by Carpenter et al. (1997, ApJ, 479, 970) in GHRS observations taken in 1992, while the $\gamma$~Cru wind seems to be very similar to that seen previously in GHRS spectra.

\begin{figure}
\plotone{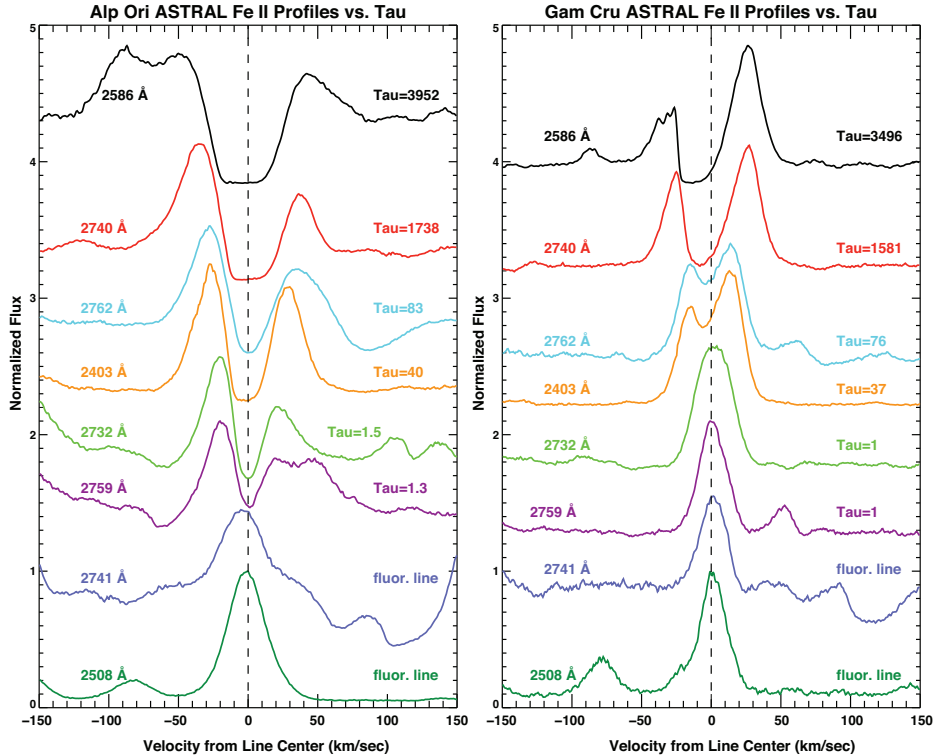}
\caption{The variation in Fe II profiles with intrinsic strength in $\alpha$~Ori (M2~Iab) and $\gamma$~Cru (M3.4~III), the two M-stars in the ASTRAL-I sample .  The shifting wavelengths of the wind-absorptions relative to the emission peaks and the changes in relative strengths of the emission peaks reflect the acceleration of the wind from the base of the chromosphere. The $\alpha$~Ori wind acceleration appears small or non-existent compared to that seen by Carpenter et al. (1997, ApJ, 479, 970) in GHRS observations taken in 1992.}
\end{figure}

\subsection{Science Example \# 2: Searching for Exotic Species in the Spectrum of the Ap Star HR~465 - Hg II, Os II, Pt II, Bi II, Tl II }

The ASTRAL-II program includes observations of the Ap star HR 465, which was chosen as a prototypical example of an A-type magnetic CP star. This spectrum is of extremely high quality in terms of both signal/noise and resolution and reveals a bright continuum that is absolutely blanketed by a sea of absorption features - so much so, that it is doubtful that a true continuum point exists anywhere in the 1200 - 3200 \AA\ range. 

HR~465 has a global magnetic field $\sim$2200 Gauss, which varies  with a period of $\sim$23.3 years \citep{Scholz_1983}. It has a spectral type of B9pe (SIMBAD), and a T$_{eff}$~$\sim$10,723~K \citep{Aller_1972} and log(g) =4.01 \citep{North_1998}.  The strength of the Cr and Eu features are also known to vary with a $\sim$23 year period \citep{Preston_1970}. \citep{Fuhrmann_1989} analyzed IUE spectra and found strong iron-peak element lines but an underabundance on some ions of low atomic number, such as C. Pt~II was seen in 1970’s IUE data, but Au~II, Hg~II, and Bi~II were uncertain and rare earth elements appeared weaker than in optical spectra.   

We have begun determinations of element abundances, concentrating on the heavy (exotic) elements of most interest, in the ASTRAL-II spectrum of HR~465, which was taken during Rare Earth Element (REE) minimum state. Our abundance measurements are obtained by comparison of the observed spectra with synthetic spectra, as shown in Figures 6 and 7.

\begin{figure}
\plotone{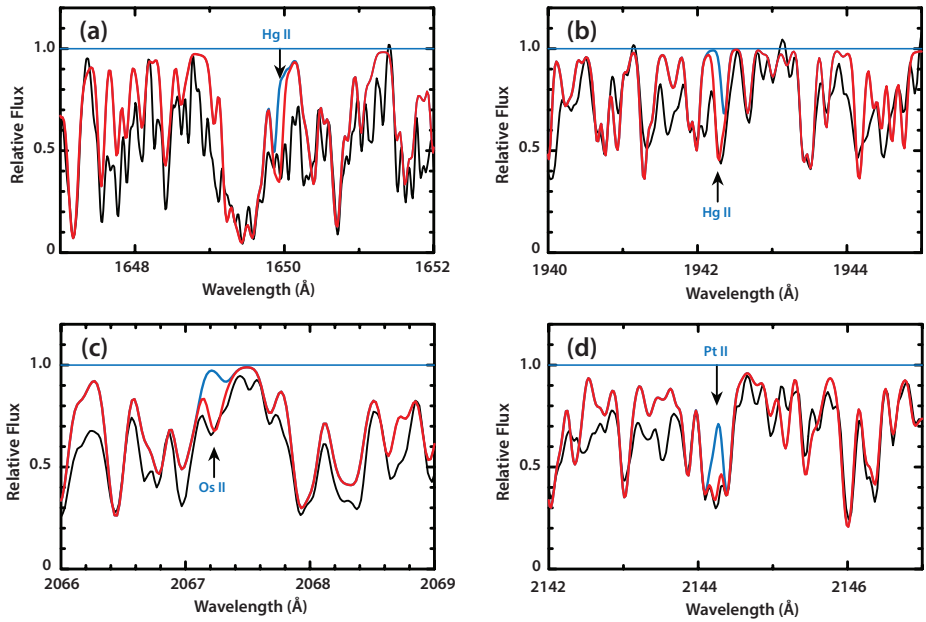}
\caption{Comparison of the observed STIS spectrum (black) with synthetic spectra computed by C. Cowley assuming T = 9900K, log g = 4.2. Plots (a) and (b) show the STIS spectrum in black and synthesized spectra in red (Hg up by 2.6 dex) and in blue (no Hg at all). The Hg II lines of interest are at 1649.95 \AA\ and 1942.27 \AA. Plot (c) is the region of the strongest Os II line, at 2067.21 \AA, with the red spectrum having Os enhanced by 3.0 dex and the blue with no Os at all. Plot (d) shows the region of a Pt line enhanced by 3.0 dex (red) and with no Pt at all (blue).}
\end{figure}

\begin{figure}
\plotone{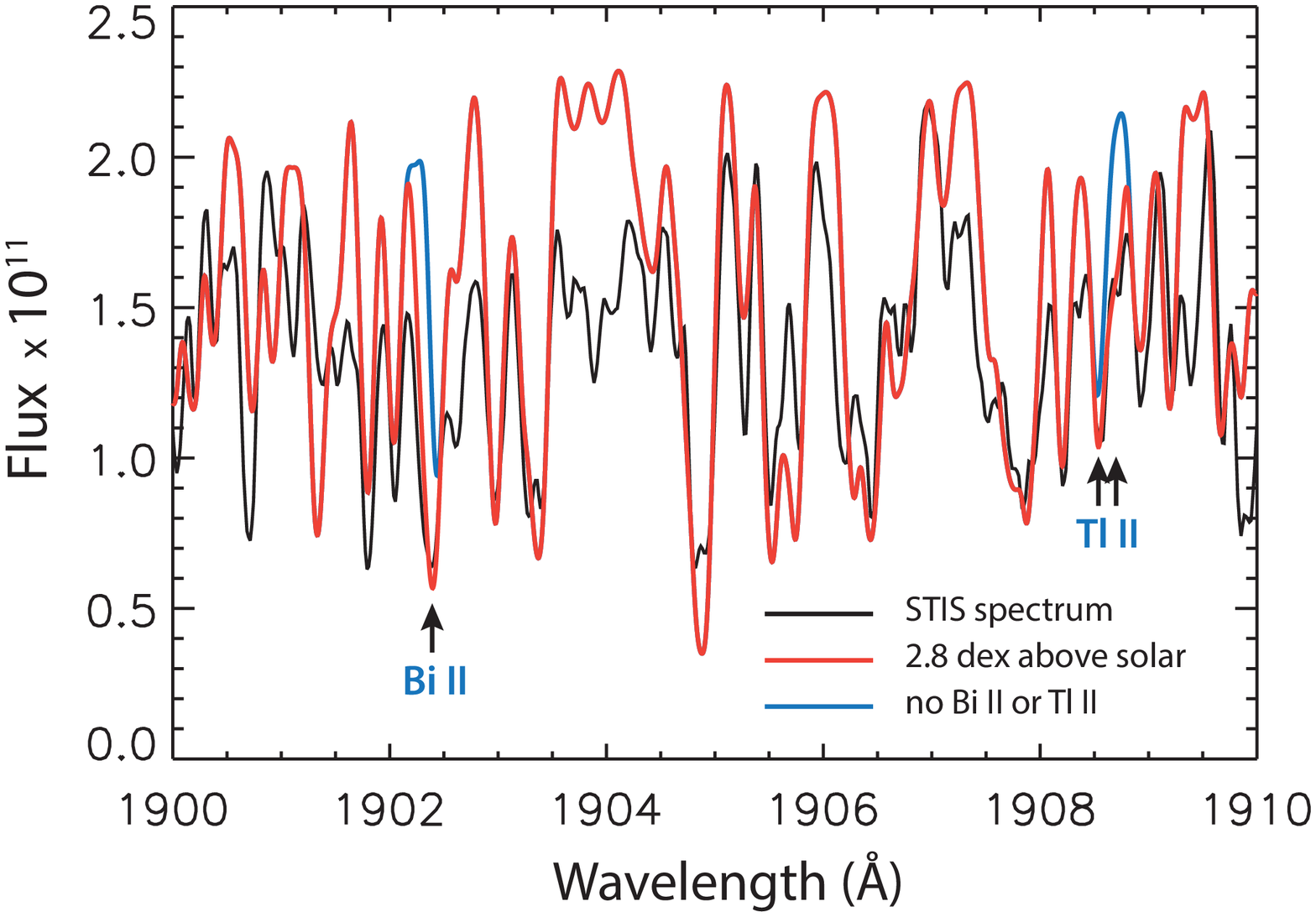}
\caption{Comparison of synthetic spectrum by G. Wahlgren with observed to search for lines of Bi II and Tl II, which do appear to be clearly present at an 2.8 dex enhancement above solar. Model uses T$_{eff}$ = 10,750~K, log g = 4.0, [M/H]=+0.5, vsini = 6.0~km/s, and v$_{turb}$ = 2.0 km/s.}
\end{figure}

%
%


\acknowledgments{Based on observations with the NASA/ESA Hubble Space Telescope obtained at the Space Telescope Science Institute, which is operated by the Association of Universities for Research in Astronomy, Incorporated, under NASA contract NAS5-26555.  This work has been supported in part by Guest Observer grants from STScI, including HST-GO-12278.05-A and HST-GO-13346.05-A.}

\end{document}